\documentclass[acmsmall]{acmart}

\usepackage{subcaption}
\usepackage{amsmath}
\usepackage{hyperref}
\usepackage{cleveref}
\usepackage{alltt}
\usepackage{tikz}
\usepackage{float}
\floatstyle{ruled}
\restylefloat{figure}

\newcommand{\racket}{\texttt{Racket}}

\newcommand{\fsm}{\texttt{FSM}}
\newcommand{\dfa}{\texttt{DFA}}
\newcommand{\ndfa}{\texttt{NDFA}}
\newcommand{\sig}{\texttt{\(\Sigma\)}}

\newcommand{\delt}{\texttt{\(\delta\)}}
\newcommand{\lamb}{\texttt{$\lambda$}}
\newcommand{\quot}{\texttt{\textquotesingle{}}}
\newcommand{\dotss}{\(\ldots\)}
\newcommand{\step}{\texttt{$\vdash$}}

\newcommand{\elist}{\texttt{\textquotesingle{()}}}
\newcommand{\qquot}{\texttt{\textasciigrave{}}}

\newcommand{\arrow}{\(\rightarrow\)}

\AtBeginDocument{%
  }

\begin{document}

\title{Visualizing Why Nondeterministic Finite-State Automata Reject}


\author{Oliwia Kempinski}
\affiliation{\institution{Seton Hall University}
             \city{South Orange}
             \state{NJ}
             \country{USA}
             \postcode{07079}}
\email{oliwia1406@gmail.com}

\author{Marco T. Moraz\'{a}n}
\affiliation{%
  \institution{Seton Hall University}
  \city{South Orange}
  \state{NJ}
  \country{USA}}
\email{morazanm@shu.edu}

\renewcommand{\shortauthors}{Moraz\'{a}n and Kempinski}

\begin{abstract}
Students find their first course in Formal Languages and Automata Theory challenging. In addition to the development of formal arguments, most students struggle to understand nondeterministic computation models. In part, the struggle stems from the course exposing them for the first time to nondeterminism. Often, students find it difficult to understand why a nondeterministic machine accepts or rejects a word. Furthermore, they may feel uncomfortable with there being multiple computations on the same input and with a machine not consuming all of its input. This article describes a visualization tool developed to help students understand nondeterministic behavior. The tool is integrated into, \fsm{}, a domain-specific language for the Automata Theory classroom. The strategy is based on the automatic generation of computation graphs given a machine and an input word. Unlike previous visualization tools, the computation graphs generated reflect the structure of the given machine's transition relation and not the structure of the computation tree.
\end{abstract}







\maketitle

\section{Introduction}

In every Formal Languages and Automata Theory course, students are asked to design nondeterministic finite-state automata (\ndfa{}). More often than not, students find this very challenging for several reasons. The first is that they are inexperienced with nondeterministic design. Second, in all likelihood, students lack the benefit of a programming language in which to implement their machines and receive immediate feedback. Such feedback is essential for students to debug a machine much like they debug a program \cite{Landa,Marwan,Venables}. Third, most machine visualization tools are ill-equipped to help students understand why a word is rejected (or accepted) by a nondeterministic automaton. In addition, students need help understanding how a machine may perform multiple computations on the same input and how, unlike deterministic machines, there can be computations that do not consume all of the input word.

In the age of domain-specific languages \cite{Bettini,Kleppe,PPL,Walter}, it is an opportunity to provide students with a programming language that makes it relatively easy to program \ndfa{}s. To this end, \fsm{} (\textbf{F}unctional \textbf{S}tate \textbf{M}achines) has been developed. \fsm{} is a domain-specific language for the Automata Theory and Formal Languages classroom embedded in \racket. Its types include \ndfa{}. Using \fsm{}, students may test and debug their machines before attempting a proof or submitting a machine for grading. Furthermore, students can implement the algorithms they develop as part of their constructive proofs. An important feature of \fsm{} is that programmers are not burdened with implementing nondeterminism. Instead, they may use nondeterminism just like they use any feature in their favorite programming language. Students design, implement, validate, and prove algorithms correct assuming nondeterminism is a built-in language feature.

One of the challenges with implementing such a language is providing support to understand and debug a nondeterministic machine addressing the problems outlined above. To understand the difficulties, it is important to recall when a nondeterministic machine accepts or rejects a given word. A nondeterministic machine accepts a given word if there is computation that takes the machine from its starting configuration to an accepting configuration. For an \ndfa{}, an accepting configuration is one in which there is no more input to consume and the machine's state is a final/accepting state. A word is rejected if there is no computation that takes the machine from its starting configuration to an accepting configuration. In other words, all possible computations must fail to end in an accepting configuration. If any possible computation ends in an accepting configuration then the machine accepts the word.

Providing support to demonstrate why a nondeterministic machine accepts a word, for example, is relatively easy. The programmer is provided with a trace of the configurations that take the machine from the starting state to a final state: C$_{\textrm{0}}$ \step{} C$_{\textrm{1}}$ \step{} \dotss C$_{\textrm{n}}$. In such a sequence, C$_{\textrm{i}}$ \step{} C$_{\textrm{j}}$ represents the application of a single rule, C$_{\textrm{0}}$ is the starting configuration, and C$_{\textrm{n}}$ is an accepting configuration. It is more challenging to provide support when a word is rejected, because for a nondeterministic machine we must show that all possible computations reject. The number of possible computations easily and quickly becomes unwieldy. Thus, providing a trace of all possible computations is of little use other than for the simplest nondeterministic machines. For this reason, until recently, \fsm{} only informed the programmer that a word is rejected. This left students wondering why the machine they designed rejected. To address this shortcoming, \fsm{} now provides the generation of computation graphs for \ndfa{}s. A computation graph is a finite visualization to help explain nondeterministic behavior.

The article is organized as follows. \Cref{rw} discusses and contrasts related work on visualization techniques for nondeterministic \ndfa{}s. \Cref{fsm} presents a brief introduction to \fsm{}. \Cref{cgraphs} presents and discusses computation graphs in \fsm{}. \Cref{ndfa-cg} discusses how \ndfa{} computation graphs are generated. Finally, \Cref{concls} presents concluding remarks and directions for future work.

\section{Related Work}
\label{rw}

In most Formal Languages and Automata Theory textbooks, computations performed by a nondeterministic machine are visualized as a computation tree (e.g., \cite{Gurari,Hopcroft,Lewis,Linz,Martin,Rich,Sipser}). In such trees, a node represents a machine configuration and an edge represents a transition. An \ndfa{} configuration is a state and the unconsumed input. Nondeterministic decisions are represented by a node having multiple children reached reading the same element or having a child with the same unconsumed input (i.e., a transition without consuming input). Computation trees provide a trace of every possible configuration a machine may reach as all possible computations progress. A configuration, therefore, may appear multiple times in a computation tree. As a consequence, there are two potential hazards for students. The first is that the tree may become unwieldy large making it difficult to understand. The second is that students are easily tempted to look for differences between two identical configurations simply because they occur in different parts of the tree.

\begin{figure}[t!]
\centering
\includegraphics[scale=0.37]{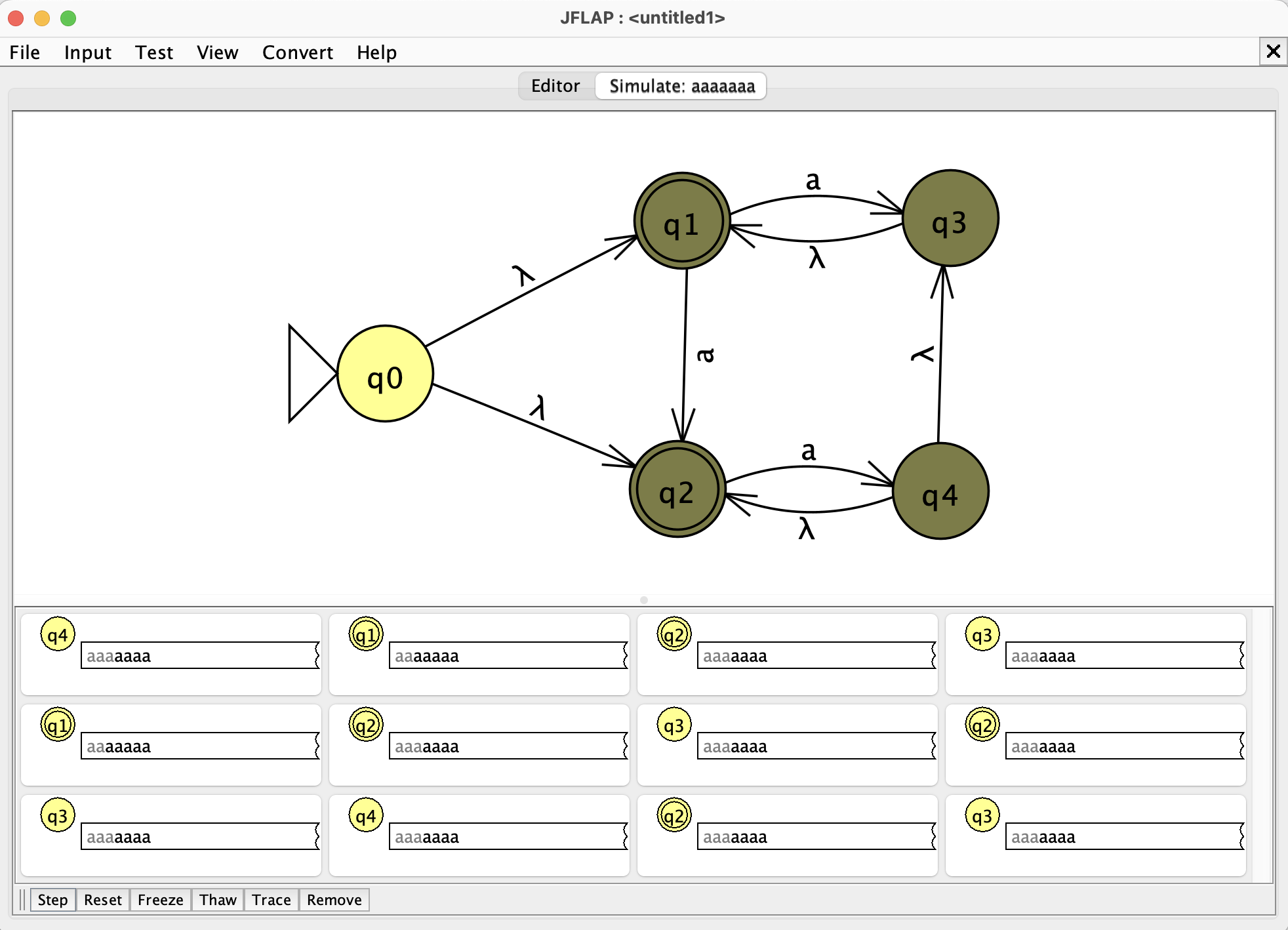}
\caption{An \ndfa{} Trace in \texttt{JFlap}.}
\label{ndfatrace}
\end{figure}

Classical automated tools to visualize why an \ndfa{} rejects a word rely on exhaustively tracing all possible configurations a machine may reach when processing a given word. That is, they trace every path in a computation tree. \texttt{JFLAP} \cite{Rodger,RodgerII} and \texttt{OpenFLAP} \cite{Mohammed}, for example, provide a trace of every possible computation. \Cref{ndfatrace} displays an \ndfa{} trace after 6 steps in \texttt{JFLAP}. Although a great deal of information is conveyed, there are two salient features that may present obstacles for learning. The first is that it is difficult to determine how a machine reached any of the configurations. The second is that the number of configurations that must be displayed can grow unwieldy and, therefore, become difficult to mentally organize. Another visualization tool is \texttt{jFAST} \cite{White}. Similar to \texttt{JFLAP}, \texttt{jFAST} also has users manually render graph-based \ndfa{} representations. Users can execute a machine by providing an input word. Unlike \texttt{JFLAP}, \texttt{jFAST} does not provide an option to see the series of transitions made during machine execution nor a trace of said transitions.

\begin{figure*}[t]
\centering
\begin{subfigure}[b]{0.49\textwidth}
\centering
\includegraphics[scale=0.6]{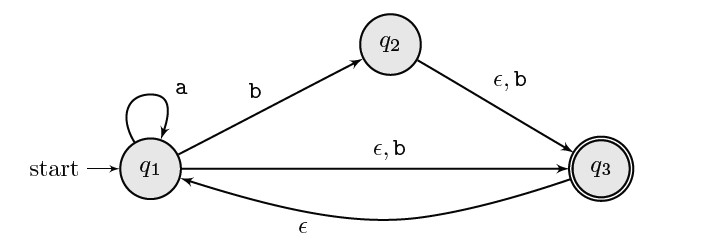}
\caption{A sample \ndfa{}.} \label{gidayu-diagram}
\end{subfigure}
\hfill
\begin{subfigure}[b]{0.49\textwidth}
\centering
\includegraphics[scale=0.6]{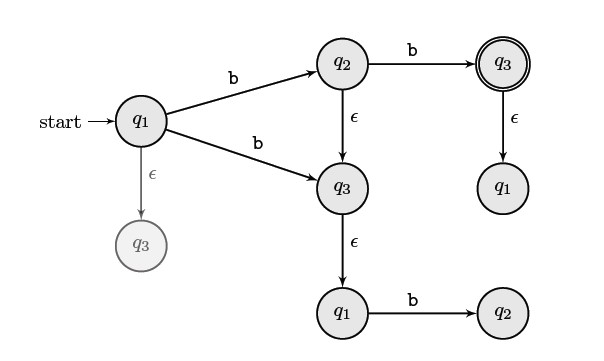}
\caption{The computation graph.} \label{gidayu-cgraph}
\end{subfigure}
\caption{A sample \ndfa{} and the \texttt{Gidayu} computation graph generated on \texttt{bb}.} \label{gdiagrams}
\end{figure*}

Modern approaches to automatically visualize \ndfa{} computations present users with a computation graph. Like a computation tree, a computation graph has nodes that represent configurations. In a computation graph, however, a configuration is represented by a single node. Thus, different computations may have shared nodes. An undeniable benefit of the approach is that the size of the visualization is tamed. \texttt{Gidayu}, for example, uses computation graphs to visualize \ndfa{} computations \cite{Gidayu}. In a \texttt{Gidayu} \ndfa{} computation graph, each node is rendered only using the configuration's state. The unconsumed input is suppressed given that it may be reconstructed by tracing back to the starting configuration's node. Thus, a machine state may appear repeatedly when it is part of different machine configurations. For instance, consider the sample \ndfa{} displayed in \Cref{gidayu-diagram} \cite[p. 111]{Gidayu} used to describe \texttt{Gidayu}. The computation graph generated for \texttt{bb} is displayed in \Cref{gidayu-cgraph} \cite[p. 112]{Gidayu}. Observe that it has three nodes labeled \texttt{q$_{\texttt{3}}$}. These nodes represent the configurations (from left to right): \texttt{(q$_{\texttt{3}}$ bb)}, \texttt{(q$_{\texttt{3}}$ b)}, and \texttt{(q$_{\texttt{3}}$ $\epsilon$)}. \texttt{Gidayu}'s computation graphs are useful for students trying to understand nondeterminism, because it is possible to determine how a configuration is reached and to determine if no accepting configuration is reached. A troublesome characteristic for some students is that a state may appear repeatedly in a computation graph. This is troublesome because students naturally associate a node with a state and not with a configuration. Another troublesome characteristic for some students is that a computation graph with nodes representing configurations does not derive its structure from the machine's state transition diagram. For instance, the computation graph in \Cref{gidayu-cgraph} is not contained in the state transition diagram displayed in \Cref{gidayu-diagram}. For many students, therefore, it is not immediately obvious how a computation graph is related to the machine's state transition diagram. Finally, there is no explicit indication that some computations do not consume the input word. For example, some students wonder what happens when the machine reaches the node representing \texttt{(q$_{\texttt{3}}$ bb)} in \Cref{gidayu-cgraph} (i.e., the leftmost occurrence of \texttt{q$_{\texttt{3}}$}). To the readers of this article, it is clear that the machine halts and rejects on that computation. When students are first exposed to nondeterministic behavior, however, it is helpful to have a computation graph that explicitly indicates that computations may end in a node without consuming all the input.

Like \texttt{Gidayu}, \fsm{} uses computation graphs, albeit different, to visualize \ndfa{} computations. In contrast to the approaches outlined above, the \fsm{} visualization reflects the structure of the machine's state transition diagram. That is, \fsm{} computation graphs are rendered based on the states traversed, not the configurations reached, by any potential computation. In this manner, the size of the visualization is bounded by the number of states and the size of the transition relation. Other than the addition of a dead state, to illustrate that some computations do not consume the entire input word, a computation graph is always a subgraph of the machine’s state diagram. Thus, students can easily see the connection between the machine’s state diagram and the visualization provided by \fsm{}. Unlike the related work discussed above, the goal is not to trace every possible computation based on the configurations traversed. Instead, students are provided with a graphic that summarizes the behavior of all possible computations. To make it easy to determine if a word is accepted or rejected, state coloring is used to highlight all the states reached at the end of every computation when the input word is consumed. If none of the colored nodes reached is a final state then the machine rejects. If at least one of the colored nodes reached is a final state then the machine accepts. Finally, to highlight that some computations do not consume all the input, a dashed edge is added to a fresh dead state from a node in which a computation halts without completely consuming the input word. In the dead state, the rest of the input is consumed. Therefore, in an \fsm{} computation graph the entire input word is consumed by every computation and students' concerns about not consuming the entire input are eased. Nonetheless, as their understanding matures, students are able to understand, that dashed edges are not part of the machine and highlight states in which computations end without consuming the entire input word.

\section{A Brief Introduction to \fsm}
\label{fsm}

\subsection{Core Definitions}

In \fsm{}, a state is an uppercase Roman alphabet symbol and an input alphabet element is a lowercase Roman alphabet symbol or a number. Based on these, an input alphabet, \texttt{$\Sigma$}, is represented as a list of alphabet symbols. A word is either \texttt{EMP}, denoting the empty word, or a nonempty list of alphabet symbols. Words are given as input to a state machine to decide if the given word is in the machine's language.

The machine constructors of interest for this article are those for deterministic and nondeterministic finite-state automata:
\begin{alltt}
     make-dfa:  K \sig{} s F \delt{} [\quot{}no-dead] \arrow{} dfa
     make-ndfa: K \sig{} s F \delt{}            \arrow{} ndfa
\end{alltt}
Here, \texttt{K} is a list of states, \texttt{F} is a list of final states in \texttt{K}, \texttt{s} is the starting state in \texttt{K}, and \texttt{$\delta$} is a transition relation. A transition relation is represented as a list of transition rules. This relation must be a function for a \dfa{}. A \dfa{} transition rule is a triple, \texttt{(K $\Sigma$ K)}, containing a source state, the element to read, and a destination state. An \ndfa{} transition rule is a triple, \texttt{(K $\{\Sigma \cup \{EMP\}\}$ K)}, containing a source state, the element to read (possibly none), and a destination state. The optional \texttt{\quot{}no-dead} argument for \texttt{make-dfa} indicates to the constructor that the relation given is a total function. Omitting this argument indicates that the transition function is incomplete and the constructor adds a fresh dead state along with transitions to this dead state for any missing transitions.

The observers are:
\begin{alltt}
     (sm-states M) (sm-sigma M) (sm-start M) (sm-finals M) (sm-rules M)
     (sm-type M)
     (sm-apply M w)
     (sm-showtransitons M w)
\end{alltt}
The first 5 observers extract a component from the given state machine. The given state machine's type is returned by \texttt{(sm-type M)}. To apply a given machine to the given word \texttt{(sm-apply M w)} is used. It returns \texttt{\quot{}accept} or \texttt{\quot{}reject}. A trace of the configurations traversed by applying a given machine to a given word is obtained using \texttt{(sm-showtransitons M w)}. A trace is only returned, however, if the machine is a \dfa{} or if the word is accepted by an \ndfa{}. For the latter, a single trace (of potentially many traces) is returned, thus, illustrating why a nondeterministic machine accepts a word. No trace is provided when an \ndfa{} rejects a word.

\subsection{Visualization}

\fsm{} provides machine rendering and machine execution visualization. The current visualization primitives are:
\begin{alltt}
     \texttt{(sm-graph M)}     \texttt{(sm-visualize M [(s p)\(\sp{*}\)])}
\end{alltt}
The first returns an image of the given machine's transition diagram. The second launches the \fsm{} visualization tool. The optional lists of length 2 contain a state of the given machine and an invariant predicate for the state. Machine execution may always be visualized if the machine is a \dfa{}. Similarly to \texttt{sm-showtransitons}, \ndfa{} machine execution may only be visualized if the given word is in the machine's language. For further details on machine execution visualization in \fsm{}, the reader is referred to a previous publication \cite{Mor8}.

\subsection{An Illustrative Example}
\label{init-ex}

To illustrate programming in \fsm{}, consider the following example:\newline
\begin{alltt}
     ;; L(M) = ab\(\sp{*}\)
     (define ab* (make-dfa \quot{}(S F)
                           \quot{}(a b)
                           \quot{}S
                           \quot{}(F)
                           \quot{}((S a F) (F b F))))

     ;; Tests
     (check-equal? (sm-apply ab* \elist) \quot{}reject)
     (check-equal? (sm-apply ab* \quot{}(b)) \quot{}reject)
     (check-equal? (sm-apply ab* \quot{}(a)) \quot{}accept)
     (check-equal? (sm-apply ab* \quot{}(a b b)) \quot{}accept)
\end{alltt}
\begin{figure*}[t!]
\centering
     \begin{subfigure}[b]{\textwidth}
         \centering
  	\includegraphics[scale=0.4]{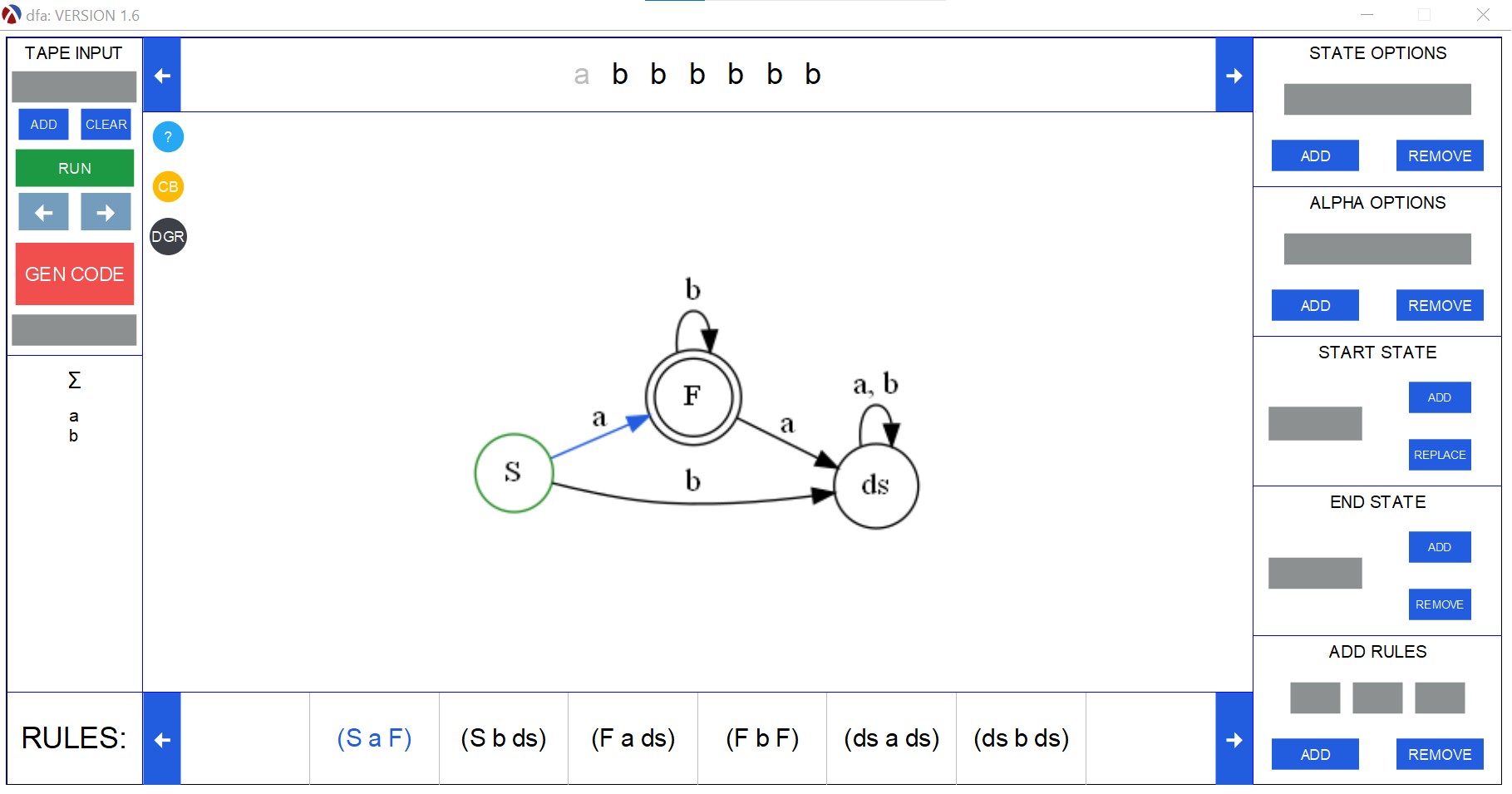}
         \caption{Transition diagram view.}
         \label{Mviz2}
     \end{subfigure}
     \hfill
     \begin{subfigure}[b]{\textwidth}
         \centering
  	\includegraphics[scale=0.3]{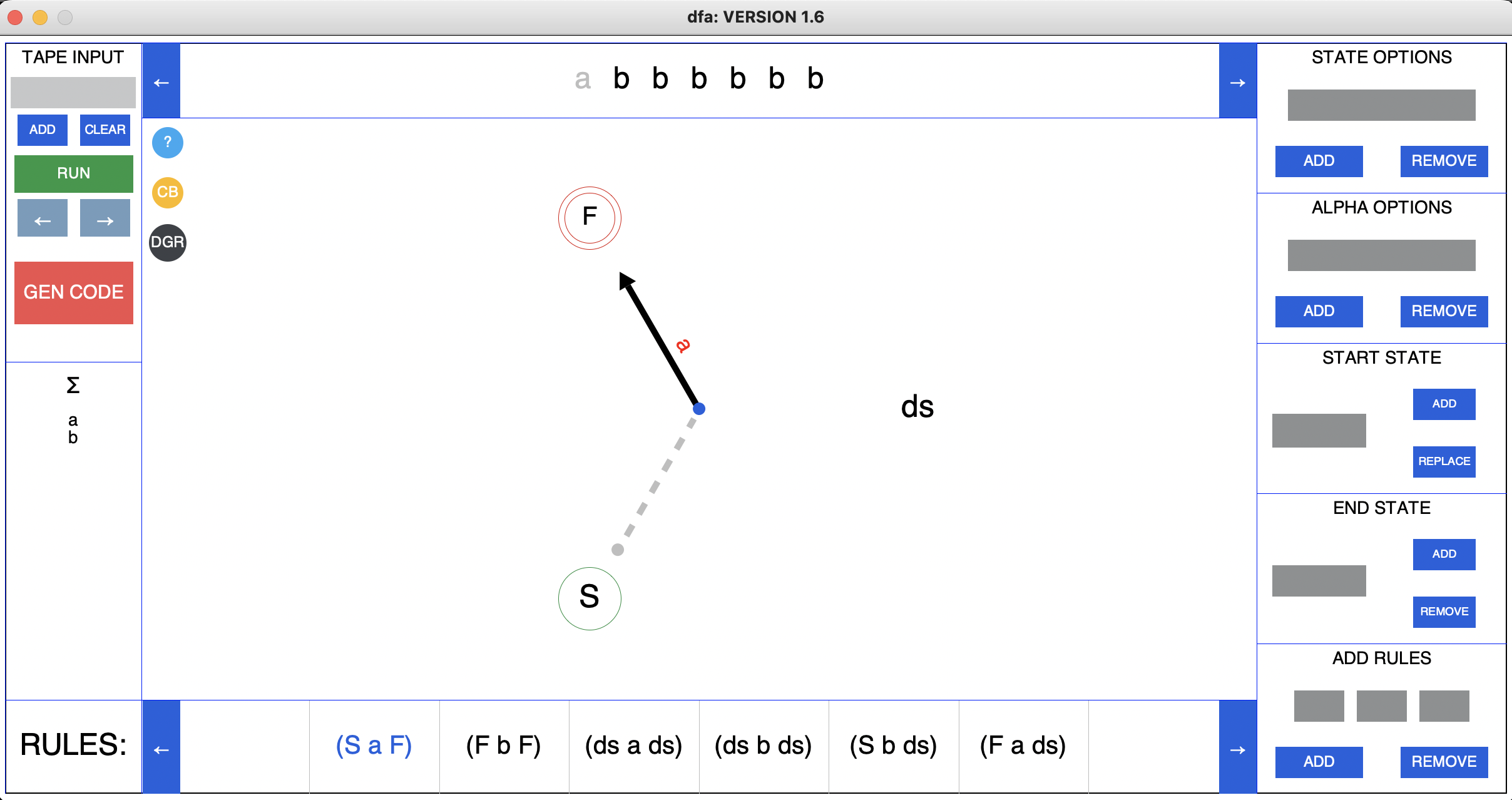}
         \caption{Control view.}
         \label{Mviz3}
     \end{subfigure}
\caption{Execution Visualization of ab* after one step.} \label{Mviz4}
\end{figure*}
The language of this machine is the set of all words that start with an \texttt{a} and end with an arbitrary number of \texttt{b}s. The constructor adds a dead state and the omitted transitions to the dead state to create a transition function. The tests validate \texttt{ab*}. Machine traces are:
\begin{alltt}
     > (sm-showtransitions ab* \quot{}(a b))
     \quot{}(((a b) S) ((b) F) (() F) accept)
     > (sm-showtransitions ab* \quot{}(b a a))
     \quot{}(((b a a) S) ((a a) ds) ((a) ds) (() ds) reject)
\end{alltt}
The trace is a list of machine configurations ending with the result. Each configuration contains the unconsumed input and the machine's state. The transition diagram obtained using \texttt{(sm-graph ab*)} is:
\begin{center}
\includegraphics[scale=0.6]{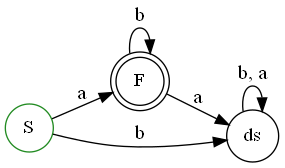}
\end{center}
The starting state is denoted in a green circle. Final states are denoted in double black circles.

Machine execution may be visualized in two views: control view and transition diagram view. \Cref{Mviz2} displays \texttt{ab*}'s state after one step using the transition diagram view. The edge representing the last transition used is highlighted in blue. \Cref{Mviz3} displays \texttt{ab*}'s  same state using the control view. The last transition used is depicted with a dashed line from the previous state to the center and a solid arrow from the center to the current state. Visualization of \ndfa{} execution is done in the same manner.

\section{Computation Graphs in \fsm{}}
\label{cgraphs}

A primary goal for computation graphs is to visually demonstrate why a nondeterministic machine rejects a given word. They are equally useful in visually explaining why a nondeterministic machine accepts a given word. Given that, more often than not, students find computation graphs based on configurations confusing, a different representation is used in \fsm{}. In an \fsm{} computation graph, nodes represent states and edges represent transitions used by any possible computation that entirely consumes the given word. To guarantee that every possible computation entirely consumes the given word, if necessary, transitions to a fresh dead state are added. In an \fsm{} computation graph, there are no repeated states. If a computation ends in a final state then we have visual proof that the word is accepted. If no computation ends in a final state then we have visual proof that the word is rejected.

There are two immediate concerns that arise with this approach. The first is that there must exist a way to distinguish states that are only traversed on all computations from states reached at the end of any computation. In an \fsm{} computation graph, this is achieved by using color. States in which any computation ends are highlighted in crimson and all other nodes are black. Observe that a state highlighted in crimson may also be an intermediate state is some computations.

The second is how to represent a computation that does not entirely consume the given word. A nondeterministic machine, unlike a deterministic machine, may reach a non-accepting configuration in which no transition rules apply. By definition, the machine halts and rejects. This makes computation graphs difficult to read, because a state may be highlighted in crimson both if the input is or is not completely consumed. To eliminate this potential pitfall, when no transitions apply, a transition consuming the next word element to a fresh non-accepting dead state is added. In the dead state, the remaining input is consumed. In this manner, all computations end in a state in which all the input is consumed. A student may easily conclude that the input word is not in the machine's language because it is completely consumed and rejected by all computations.

\section{\ndfa{} Computation Graphs}
\label{ndfa-cg}

A nice characteristic of \dfa{}s and \ndfa{}s is that they decide a language. That is, they always accept or reject the given word. This means that a computation graph can always be constructed because every possible computation is guaranteed to end. This holds in \fsm{} even when there are loops only involving $\epsilon$-transitions, because \fsm{}'s implementation of nondeterminism does not explore the same configuration more than once.

\subsection{Design Idea}
\label{DI}

To build an \ndfa{} computation graph from a given machine and a given word, a breadth-first traversal of the computation tree is performed. A configuration is represented as a structure, \texttt{ndfa-stuci}, that is defined as follows:
\begin{alltt}
     ;; An ndfa configuration is a structure, (ndfa-stuci state word),
     ;; containing a state and the unconsumed input.
     (struct ndfa-stuci [state ui] #:transparent)
\end{alltt}
The search takes as input a \texttt{(listof ndfa-stuci)}\footnote{This denotes a type: a list that contains zero or more \texttt{ndfa-stuci}s.} for the configurations to explore and an accumulator for the explored configurations. The accumulator is used to prevent processing the same configuration more than once. For each unexplored \texttt{ndfa-stuci}, new computation graph edges and new \texttt{ndfa-stuci}s are created by using all applicable transitions. If the list of new \texttt{ndfa-stuci}s is empty then the search stops and the new list of computational graph edges is returned. Otherwise, the new list of computational graph edges is appended to the result of recursively processing the new \texttt{ndfa-stuci}s and a new accumulator that is obtained by appending the given \texttt{ndfa-stuci}s and the accumulator. In essence, at every step the next level of the computation tree is generated for exploration.

There are two varieties of edges: regular edges, \texttt{ndfa-edge}s, that have a destination state that does not need to be highlighted (i.e., not the last transition used in a computation) and \texttt{sp}ecial \texttt{edge}s, \texttt{ndfa-spedge}s, that have a destination state that needs to be highlighted (i.e., a transition used in a computation that leaves the unconsumed input empty). They are defined as follows:
\begin{alltt}
     ;; An ndfa-Edge is structure that is either:
     ;; 1. (ndfa-edge state symb state)
     ;; 2. (ndfa-spedge state symb state)
\end{alltt}
For a given \texttt{ndfa-stuci}, the new edges generated may be of either type. A conditional expression is used, outlined as follows, to generate the new edges:
\begin{description}
  \item[Unconsumed input is empty] New \texttt{ndfa-spedge}s are generated from applicable empty transitions. Only \texttt{ndfa-spedge}s are created because for any state reached the word is entirely consumed.
  \item[Length of unconsumed input is 1] This means that the given \texttt{ndfa-stuci} represents a configuration that has a single input element left to process. A list of new edges is created by appending the list of \texttt{ndfa-spedge}s obtained from applicable transition rules that consume the last input element and the list of \texttt{ndfa-edge}s obtained from applicable empty transition rules.
  \item[Length of unconsumed input $>$ 1] This means that the given \texttt{ndfa-stuci} represents a configuration that has more than a single input element left to process. A list of new \texttt{ndfa-edge}s is created using all applicable rules. No new \texttt{ndfa-spedges} are generated because none of the applicable rules empty the unconsumed input.
\end{description}
On an empty list of new edges, a list containing an \texttt{ndfa-spedge} to the fresh dead state consuming the word's next element is returned. On a new list of edges that only represent empty transitions, an \texttt{ndfa-spedge} to the fresh dead state consuming the word's next element is added to the list of new edges. It is safe to make these \texttt{ndfa-spedge}s because the rest of the input for any such computation is consumed in the fresh dead state. Otherwise, the list of new edges is returned.

The returned list of computation graph edges is an overestimation. It may contain repetitions, transitions represented as both an \texttt{ndfa-edge} and a \texttt{ndfa-spedge}, and redundant edges if the word is accepted by the given machine. Therefore, the returned list of edges is filtered to remove duplicates and to remove any \texttt{ndfa-edge}s that are also \texttt{ndfa-spedge}s. The \texttt{ndfa-edge}s, not the \texttt{ndfa-spedge}s, are removed because the \texttt{spedges} are needed to correctly highlight states reached when the word is entirely consumed. Finally, if the given word is accepted by the given machine then the computation graph ought to only contain edges that take the machine from the starting state to an accepting state consuming the given word. This is achieved using \texttt{sm-showtransitions}. From the result returned by \texttt{sm-showtransitions}, the set of edges traversed is extracted. Any computation-graph edge not in this set is discarded.

Finally, to generate the computation graph, \texttt{ndfa-edge}s are rendered as solid lines given that they correspond to edges in the machine's state transition diagram. The rendering of \texttt{ndfa-spedge}s requires examining the destination state. If the destination state is a machine state then the edge is rendered as a solid black arrow. If the destination state is the fresh dead state then the edge is rendered as a dashed black arrow. In both cases, the destination state is highlighted in crimson.

\subsection{Illustrative Example}
\label{ndfa-ex}

\begin{figure}[t!]
\centering
\includegraphics[scale=0.5]{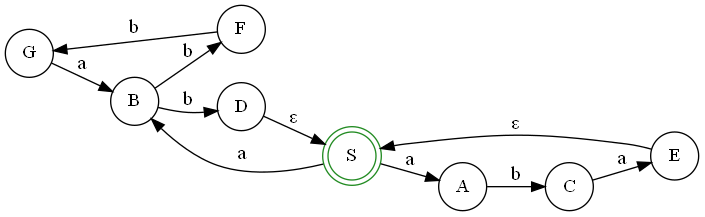}
\caption{\ndfa{} for L = ((a b a) \(\cup\) (a (b b a)\(\sp{*}\) b))\(\sp{*}\).}
\label{ndfa3}
\end{figure}

To illustrate the result of implementing the algorithm outlined, consider the following \ndfa{} designed by a student:
\begin{alltt}
     ;; L = ((a b a) \(\cup\) (a (b b a)\(\sp{*}\) b))\(\sp{*}\)
     (define M (make-ndfa \quot{}(S A B C D E F G)
                          \quot{}(a b)
                          \quot{}S
                          \quot{}(S)
                          \qquot{}((S a A) (S a B) (A b C)
                            (B b D) (B b F) (C a E)
                            (D ,EMP S) (E ,EMP S) (F b G) (G a B))))
     ;; Tests
     (check-equal? (sm-apply M \quot{}(b b)) \quot{}reject)
     (check-equal? (sm-apply M \quot{}(a b a a b b)) \quot{}reject)
     (check-equal? (sm-apply M \quot{}()) \quot{}accept)
     (check-equal? (sm-apply M \quot{}(a b a a b)) \quot{}accept)
     (check-equal? (sm-apply M \quot{}(a b a a b a)) \quot{}accept)
\end{alltt}
The state transition diagram is displayed in \Cref{ndfa3}. The language is the concatenation of an arbitrary number of \texttt{\quot{}(a b a)} and \texttt{\quot{}((a (b b a))\(\sp{*}\) b)))}.

\begin{figure}[t!]
\centering
\includegraphics[scale=0.45]{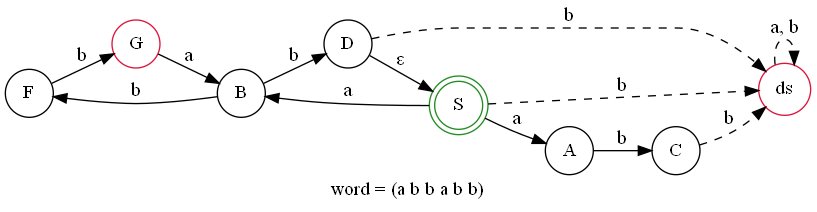}
\caption{Computation graph on reject for \ndfa{} in \Cref{ndfa3}.}
\label{ndfa3cg}
\end{figure}

The computation graph obtained from applying \texttt{M} to \texttt{word=\quot{}(a b b a b b)} is displayed in \Cref{ndfa3cg}. The only states highlighted in crimson are \texttt{G} and \texttt{ds}, which are not final states. Thus, no computations end in an accepting state and, therefore, it is straight-forward to see that the word is rejected by the machine. Also notable is that there are both computations that do not consume and computations that consume all the input. The existence of computations that do not consume all the input is extrapolated from the transitions into \texttt{ds} from \texttt{S}, \texttt{C}, and \texttt{D}. These transitions indicate that these states are reached, but the machine is unable to consume the rest of the input. The existence of computations that consume all the input are extrapolated from machine states, like \texttt{G}, that are highlighted in crimson. For such a state, the state is reached by a computation after consuming the given word but the computation rejects given that \texttt{G} is not a final state. Thus, we have a clear visualization that informs us that the word is rejected, that not all computations completely consume the word, and that some computations completely consume the word.

\begin{figure}[t!]
\centering
\includegraphics[scale=0.5]{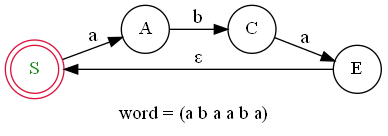}
\caption{Computation graph on accept for \ndfa{} in \Cref{ndfa3}.} \label{ndfa-cgraph-accept}
\end{figure}

Finally, it is enlightening to illustrate the computation graph generated when the given word is accepted. \Cref{ndfa-cgraph-accept} displays the \fsm{} computation graph obtained from applying \texttt{M} to the word \texttt{\quot{}(a b a a b a)}. It is easily observed that the word is accepted given that a single final state is highlighted in crimson. Furthermore, only the edges representing the transitions used in a single accepting computation are displayed. This suffices to conclude that the machine accepts the word. There is no need to explore nor display other computations, if any, that also accept the word.

\subsection{Implementation Sketch}
\label{ndfa-impl}

The function to generate computation-graph edges may be elegantly implemented using higher-order functions. This function composes the functions to traverse the computation tree, to remove duplicates, to remove redundant \texttt{ndfa-edge}s, and to remove redundant edges when the machine accepts the word as follows:
\begin{alltt}
     ;; fsa word \arrow{} (listof ndfa-Edge)
     ;; Purpose: Return computation-graph edges for given fsa and word.
     (define (make-cg-edges M word)
       ((compose remove-redundant-edges-on-accept
                 remove-duplicate-Edges
                 remove-duplicates
                 computation-tree->cg-edges)
        (list (ndfa-stuci (sm-start M) word)) \elist{}))
\end{alltt}
The arguments are the initial values for the two accumulators needed to traverse the computation tree: the only configuration to explore is the initial configuration and there are no configurations that have been explored.

\begin{figure}[t!]
\begin{alltt}
     ;; (listof ndfa-stuci) (listof ndfa-stuci) \arrow{} (listof ndfa-Edge)
     ;; Purpose: Return the list of ndfa-edges and ndfa-spedges obtained by
     ;;          traversing the computation tree using BFS
     ;; Accumulator Invariants:
     ;;  to-visit = list of unexplored ndfa-stucis
     ;;  visited  = list of explored ndfa-stucis
     (define (computation-tree->cg-edges to-visit visited)
       (let* [(new-edges
                (remove-duplicates
                  (append-map (\lamb{} (s) (add-to-edges s)) to-visit)))
              (new-stucis
                (add-to-stucis new-edges to-visit visited))]
         (if (empty? new-stucis)
             new-edges
             (append new-edges
                     (computation-tree->cg-edges new-stucis
                                                 (append to-visit visited))))))
\end{alltt}
\caption{The function to traverse the computation tree.} \label{ct-traversal}
\end{figure}

The function that performs the computation tree's breadth-first search is displayed in \Cref{ct-traversal}. The function creates new edges from the \texttt{ndfa-stuci}s that must be visited. From these edges, the next \texttt{ndfa-stuci}s needed are computed without recreating any in \texttt{to-visit} nor \texttt{visited}. If there are no new \texttt{ndfa-stuci}s to explore then the new list of edges is returned. Otherwise, the new list of edges is appended to the result of recursively exploring the new \texttt{ndfa-stuci}s and adding the given \texttt{ndfa-stuci}s to \texttt{visited}.

\begin{figure}[t!]
\begin{alltt}
;; ndfa-stuci \arrow{} (listof ndfa-Edge)
;; Purpose: Compute ndfa-Edges for given ndfa-stuci
(define (add-to-edges stuci)
  (if (empty? (ndfa-stuci-ui stuci))
      (map (\lamb{} (r) (ndfa-spedge (ndfa-rule-fromst r)
                               (ndfa-rule-read r)
                               (ndfa-rule-tost r)))
           (filter (\lamb{} (r) (and (eq? EMP (ndfa-rule-read r))
                               (eq? (ndfa-stuci-state stuci)
                                    (ndfa-rule-fromst r))))
                   (sm-rules M)))
      (let* [(new-rules
               (if (= 1 (length (ndfa-stuci-ui stuci)))
                   (append (map (\lamb{} (r) (ndfa-spedge (ndfa-rule-fromst r)
                                                    (ndfa-rule-read r)
                                                    (ndfa-rule-tost r)))
                                (filter nonempty-rules (sm-rules M)))
                           (map (\lamb{} (r) (ndfa-edge (ndfa-rule-fromst r)
                                                  (ndfa-rule-read r)
                                                  (ndfa-rule-tost r)))
                                (filter empty-rules (sm-rules M))))
                   (map (\lamb{} (r) (ndfa-edge (ndfa-rule-fromst r)
                                          (ndfa-rule-read r)
                                          (ndfa-rule-tost r)))
                        (filter empty-or-nonempty-rules (sm-rules M)))))]
              (cond [(empty? new-rules)
                     (list (ndfa-spedge (ndfa-stuci-state stuci)
                                        (first (ndfa-stuci-ui stuci))
                                        my-ds))]
                    [(andmap empty-trans? (map ndfa-Edge-read new-rules))
                     (cons (ndfa-spedge (ndfa-stuci-state stuci)
                                        (first (ndfa-stuci-ui stuci))
                                        my-ds)
                           new-rules)]
                    [else new-rules]))))
\end{alltt}
\caption{Function to compute new edges from given ndfa-stuci.} \label{newe}
\end{figure}

The function to compute new edges from an unexplored \texttt{ndfa-stuci} is displayed in \Cref{newe}. If the given \texttt{ndfa-stuci}'s unconsumed input is empty, \texttt{ndfa-spedge}s are created for all applicable empty transition rules. If the unconsumed input is not empty, for all applicable rules, new computation graph edges are created. If the length of the unconsumed input is 1 then \texttt{ndfa-spedge}s and \texttt{ndfa-edge}s are created using, respectively, the non-empty and the empty transition rules. Otherwise, only \texttt{ndfa-edges} are created using all applicable rules. As outlined in \Cref{DI}, if the new list of edges is empty then a single \texttt{ndfa-spedge} to the fresh dead state is generated. If the new list of edges is generated from only empty transitions then an \texttt{ndfa-spedge} to the fresh dead state is added to the new rules. Otherwise, the new rules are returned.

\begin{figure}[t!]
\begin{alltt}
\begin{small}
;; (listof ndfa-Edge) (listof ndfa-stuci) (listof ndfa-stuci) \arrow{} (listof ndfa-stuci)
;; Purpose: Compute new ndfa-stucis from the given ndfa-stucis to explore
;; Accumulator Invariant: visited = list of generated ndfa-stucis
(define (add-to-stucis Edges to-visit visited)
  ;; ndfa-stuci \arrow{} (listof ndfa-stuci)
  ;; Purpose: Compute new ndfa-stucis from the given ndfa-stuci
  (define (add-to-stucis-helper stuci)
    (if (empty? (ndfa-stuci-ui stuci))
        (map (\lamb{} (e) (ndfa-stuci (ndfa-Edge-tost e) (ndfa-stuci-ui stuci)))
             (filter (\lamb{} (e) (and (eq? (ndfa-Edge-read e) EMP)
                                 (eq? (ndfa-Edge-fromst e)
                                      (ndfa-stuci-state stuci))))
                     Edges))
        (map (\lamb{} (e) (ndfa-stuci (ndfa-Edge-tost e)
                                (if (eq? (ndfa-Edge-read e)
                                         (first (ndfa-stuci-ui stuci)))
                                    (rest (ndfa-stuci-ui stuci))
                                    (ndfa-stuci-ui stuci))))
             (filter (\lamb{} (e) (and (or (eq? (ndfa-Edge-read e)
                                          (first (ndfa-stuci-ui stuci)))
                                     (eq? (ndfa-Edge-read e) EMP))
                                 (eq? (ndfa-Edge-fromst e)
                                      (ndfa-stuci-state stuci))))
                     Edges))))
  (if (empty? to-visit)
      \elist{}
      (let* [(new-stucis (add-to-stucis-helper (first to-visit)))]
        (if (or (empty? new-stucis)
                (andmap (\lamb{} (s) (member s visited)) new-stucis))
            (add-to-stucis Edges (rest to-visit) visited)
            (append new-stucis
                    (add-to-stucis Edges
                                   (rest to-visit)
                                   (append new-stucis visited)))))))
\end{small}
\end{alltt}
\caption{The function to compute new \texttt{ndfa-stucis}.} \label{add2stucis}
\end{figure}

Finally, the function to create new \texttt{ndfa-stuci}s from the list of \texttt{ndfa-Edges} is presented in \Cref{add2stucis}. If the list of \texttt{ndfa-stuci}s to visit is empty, the function halts and returns an empty list of \texttt{ndfa-stuci}s. Otherwise, the given \texttt{ndfa-stuci}s are traversed to compute the new \texttt{ndfa-stuci}s at the next level of the computation tree. If there are no new \texttt{ndfa-stuci}s or all are already visited then the rest of the \texttt{ndfa-stuci}s are processed recursively. Otherwise, the new \texttt{ndfa-stuci}s are appended to the result of recursively processing the rest of the given \texttt{ndfa-stuci}s and adding the given \texttt{ndfa-stuci}s to the accumulator for generated \texttt{ndfa-stuci}s. The list of new \texttt{ndfa-stuci}s for a given \texttt{ndfa-stuci} is generated by \texttt{add-to-stucis-helper} in \Cref{add2stucis}. New \texttt{ndfa-stuci}s are generated using the applicable edges' destination states. If the given \texttt{ndfa-stuci}'s unconsumed input's first element matches the element consumed by an edge then the unconsumed input is the rest of the given \texttt{ndfa-stuci}'s unconsumed input. Otherwise, the unconsumed input is unchanged (i.e., the edge represents an empty transition).

\section{Concluding Remarks}
\label{concls}

This article presents a novel graphical approach to help students understand why nondeterministic finite-state automata reject a given word. The approach is integrated into \fsm{}--a domain-specific language that allows programmers to easily build and execute such machines. Unlike previous approaches, that focus on tracing machine configurations, the work presented here builds on the transition diagram of a given machine. The nodes in an \fsm{} computation graph represent states traversed by any computation and the edges represent the transitions used in any computation. Computation graphs are purposely built with a fresh dead state, if needed, so that all computations may consume all their input. States reached that satisfy this criteria, in any possible computation, are highlighted in crimson to indicate where at least one computation ends. The result is the generation of computation graphs that summarize what happens on all computations. They allow students to easily determine if any crimson states are final states. If none are, then they can see that the machine rejects because no computation ends in an accepting configuration. In addition, students easily associate a computation graph with a machine's state transition diagram and can easily determine transitions used and states traversed during any computation. Finally, computations for accepted words only contain states and transitions traversed to reach an accepting configuration.

Future work includes the generation of computation graphs for pushdown automata and for Turing machines that decide and semidecide a language. Generation of such graphs requires care, because the length of a computation may be infinite. This means that in some cases it may only be possible to approximate a computation graph.


\bibliographystyle{ACM-Reference-Format}
\bibliography{Computation-Graphs}


\begin{thebibliography}{20}


\ifx \showCODEN    \undefined \def \showCODEN     #1{\unskip}     \fi
\ifx \showDOI      \undefined \def \showDOI       #1{#1}\fi
\ifx \showISBNx    \undefined \def \showISBNx     #1{\unskip}     \fi
\ifx \showISBNxiii \undefined \def \showISBNxiii  #1{\unskip}     \fi
\ifx \showISSN     \undefined \def \showISSN      #1{\unskip}     \fi
\ifx \showLCCN     \undefined \def \showLCCN      #1{\unskip}     \fi
\ifx \shownote     \undefined \def \shownote      #1{#1}          \fi
\ifx \showarticletitle \undefined \def \showarticletitle #1{#1}   \fi
\ifx \showURL      \undefined \def \showURL       {\relax}        \fi
\providecommand\bibfield[2]{#2}
\providecommand\bibinfo[2]{#2}
\providecommand\natexlab[1]{#1}
\providecommand\showeprint[2][]{arXiv:#2}

\bibitem[Bettini(2016)]%
        {Bettini}
\bibfield{author}{\bibinfo{person}{L. Bettini}.}
  \bibinfo{year}{2016}\natexlab{}.
\newblock \bibinfo{booktitle}{\emph{{Implementing Domain-Specific Languages
  with Xtext and Xtend}}}.
\newblock \bibinfo{publisher}{Packt Publishing}.
\newblock
\showISBNx{9781786463272}


\bibitem[Cogumbreiro and Blike(2022)]%
        {Gidayu}
\bibfield{author}{\bibinfo{person}{Tiago Cogumbreiro} {and}
  \bibinfo{person}{Gregory Blike}.} \bibinfo{year}{2022}\natexlab{}.
\newblock \showarticletitle{{Gidayu: Visualizing Automaton and Their
  Computations}}. In \bibinfo{booktitle}{\emph{Proceedings of the 27th ACM
  Conference on on Innovation and Technology in Computer Science Education Vol.
  1}} (Dublin, Ireland) \emph{(\bibinfo{series}{ITiCSE '22})}.
  \bibinfo{publisher}{Association for Computing Machinery},
  \bibinfo{address}{New York, NY, USA}, \bibinfo{pages}{110–116}.
\newblock
\showISBNx{9781450392013}
\urldef\tempurl%
\url{https://doi.org/10.1145/3502718.3524742}
\showDOI{\tempurl}


\bibitem[Felleisen et~al\mbox{.}(2018)]%
        {PPL}
\bibfield{author}{\bibinfo{person}{Matthias Felleisen},
  \bibinfo{person}{Robert~Bruce Findler}, \bibinfo{person}{Matthew Flatt},
  \bibinfo{person}{Shriram Krishnamurthi}, \bibinfo{person}{Eli Barsilay},
  \bibinfo{person}{Jay McCarthy}, {and} \bibinfo{person}{Sam Tobin-Hochstadt}.}
  \bibinfo{year}{2018}\natexlab{}.
\newblock \showarticletitle{{A Programmable Programming Language}}.
\newblock \bibinfo{journal}{\emph{Commun. ACM}} \bibinfo{volume}{61},
  \bibinfo{number}{13} (\bibinfo{date}{March} \bibinfo{year}{2018}),
  \bibinfo{pages}{62--71}.
\newblock
\showISSN{0001-0782}
\urldef\tempurl%
\url{https://doi.org/10.1145/3127223}
\showDOI{\tempurl}


\bibitem[Gurari(1989)]%
        {Gurari}
\bibfield{author}{\bibinfo{person}{Eitan~M. Gurari}.}
  \bibinfo{year}{1989}\natexlab{}.
\newblock \bibinfo{booktitle}{\emph{{An Introduction to the Theory of
  Computation}}}.
\newblock \bibinfo{publisher}{Computer Science Press}.
\newblock
\showISBNx{0-7167-8182-4}


\bibitem[Hopcroft et~al\mbox{.}(2006)]%
        {Hopcroft}
\bibfield{author}{\bibinfo{person}{John~E. Hopcroft}, \bibinfo{person}{Rajeev
  Motwani}, {and} \bibinfo{person}{Jeffrey~D. Ullman}.}
  \bibinfo{year}{2006}\natexlab{}.
\newblock \bibinfo{booktitle}{\emph{{Introduction to Automata Theory,
  Languages, and Computation (3rd Edition)}}}.
\newblock \bibinfo{publisher}{Addison-Wesley Longman Publishing Co., Inc.},
  \bibinfo{address}{USA}.
\newblock
\showISBNx{0321455363}


\bibitem[Kleppe(2008)]%
        {Kleppe}
\bibfield{author}{\bibinfo{person}{A. Kleppe}.}
  \bibinfo{year}{2008}\natexlab{}.
\newblock \bibinfo{booktitle}{\emph{{Software Language Engineering: Creating
  Domain-Specific Languages Using Metamodels}}}.
\newblock \bibinfo{publisher}{Pearson Education}.
\newblock
\showISBNx{9780321606464}


\bibitem[Landa and Martinez-Trevino(2019)]%
        {Landa}
\bibfield{author}{\bibinfo{person}{Raquel Landa} {and} \bibinfo{person}{Yolanda
  Martinez-Trevino}.} \bibinfo{year}{2019}\natexlab{}.
\newblock \showarticletitle{Relevance of Immediate Feedback in an Introduction
  to Programming Course}. In \bibinfo{booktitle}{\emph{2019 ASEE Annual
  Conference \& Exposition}}. \bibinfo{publisher}{ASEE Conferences},
  \bibinfo{address}{Tampa, Florida}.
\newblock
\urldef\tempurl%
\url{https://doi.org/10.18260/1-2--33235}
\showDOI{\tempurl}


\bibitem[Lewis and Papadimitriou(1997)]%
        {Lewis}
\bibfield{author}{\bibinfo{person}{Harry~R. Lewis} {and}
  \bibinfo{person}{Christos~H. Papadimitriou}.}
  \bibinfo{year}{1997}\natexlab{}.
\newblock \bibinfo{booktitle}{\emph{Elements of the Theory of Computation}
  (\bibinfo{edition}{2nd} ed.)}.
\newblock \bibinfo{publisher}{Prentice Hall PTR}, \bibinfo{address}{Upper
  Saddle River, NJ, USA}.
\newblock
\showISBNx{0132624788}
\urldef\tempurl%
\url{https://doi.org/10.1145/300307.1040360}
\showDOI{\tempurl}


\bibitem[Linz(2011)]%
        {Linz}
\bibfield{author}{\bibinfo{person}{Peter Linz}.}
  \bibinfo{year}{2011}\natexlab{}.
\newblock \bibinfo{booktitle}{\emph{{An Introduction to Formal Languages and
  Automata}} (\bibinfo{edition}{5th} ed.)}.
\newblock \bibinfo{publisher}{Jones and Bartlett Publishers, Inc.},
  \bibinfo{address}{USA}.
\newblock
\showISBNx{9781449615529}


\bibitem[Martin(2003)]%
        {Martin}
\bibfield{author}{\bibinfo{person}{John~C. Martin}.}
  \bibinfo{year}{2003}\natexlab{}.
\newblock \bibinfo{booktitle}{\emph{Introduction to Languages and the Theory of
  Computation} (\bibinfo{edition}{3} ed.)}.
\newblock \bibinfo{publisher}{McGraw-Hill, Inc.}, \bibinfo{address}{New York,
  NY, USA}.
\newblock
\showISBNx{0072322004, 9780072322002}


\bibitem[Marwan et~al\mbox{.}(2020)]%
        {Marwan}
\bibfield{author}{\bibinfo{person}{Samiha Marwan}, \bibinfo{person}{Ge Gao},
  \bibinfo{person}{Susan Fisk}, \bibinfo{person}{Thomas~W. Price}, {and}
  \bibinfo{person}{Tiffany Barnes}.} \bibinfo{year}{2020}\natexlab{}.
\newblock \showarticletitle{{Adaptive Immediate Feedback Can Improve Novice
  Programming Engagement and Intention to Persist in Computer Science}}. In
  \bibinfo{booktitle}{\emph{Proceedings of the 2020 ACM Conference on
  International Computing Education Research}} \emph{(\bibinfo{series}{ICER
  '20})}. \bibinfo{publisher}{Association for Computing Machinery},
  \bibinfo{address}{New York, NY, USA}, \bibinfo{pages}{194–203}.
\newblock
\showISBNx{9781450370929}
\urldef\tempurl%
\url{https://doi.org/10.1145/3372782.3406264}
\showDOI{\tempurl}


\bibitem[Mohammed(2020)]%
        {Mohammed}
\bibfield{author}{\bibinfo{person}{Mostafa Kamel~Osman Mohammed}.}
  \bibinfo{year}{2020}\natexlab{}.
\newblock \showarticletitle{Teaching Formal Languages through Visualizations,
  Simulators, Auto-graded Exercises, and Programmed Instruction}. In
  \bibinfo{booktitle}{\emph{Proceedings of the 51st {ACM} Technical Symposium
  on Computer Science Education, {SIGCSE} 2020, Portland, OR, USA, March 11-14,
  2020}}, \bibfield{editor}{\bibinfo{person}{Jian Zhang}, \bibinfo{person}{Mark
  Sherriff}, \bibinfo{person}{Sarah Heckman}, \bibinfo{person}{Pamela~A.
  Cutter}, {and} \bibinfo{person}{Alvaro~E. Monge}} (Eds.).
  \bibinfo{publisher}{{ACM}}, \bibinfo{pages}{1429}.
\newblock
\urldef\tempurl%
\url{https://doi.org/10.1145/3328778.3372711}
\showDOI{\tempurl}


\bibitem[Moraz{\'{a}}n et~al\mbox{.}(2020)]%
        {Mor8}
\bibfield{author}{\bibinfo{person}{Marco~T. Moraz{\'{a}}n},
  \bibinfo{person}{Joshua~M. Schappel}, {and} \bibinfo{person}{Sachin
  Mahashabde}.} \bibinfo{year}{2020}\natexlab{}.
\newblock \showarticletitle{Visual Designing and Debugging of Deterministic
  Finite-State Machines in {FSM}}.
\newblock \bibinfo{journal}{\emph{Electronic Proceedings in Theoretical
  Computer Science}}  \bibinfo{volume}{321} (\bibinfo{date}{aug}
  \bibinfo{year}{2020}), \bibinfo{pages}{55--77}.
\newblock
\urldef\tempurl%
\url{https://doi.org/10.4204/eptcs.321.4}
\showDOI{\tempurl}


\bibitem[Rich(2019)]%
        {Rich}
\bibfield{author}{\bibinfo{person}{Elaine Rich}.}
  \bibinfo{year}{2019}\natexlab{}.
\newblock \bibinfo{booktitle}{\emph{Automata, Computability and Complexity:
  Theory and Applications}}.
\newblock \bibinfo{publisher}{Pearson Prentice Hall}.
\newblock
\showISBNx{9780132288064}


\bibitem[Rodger(2006)]%
        {Rodger}
\bibfield{author}{\bibinfo{person}{Susan~H. Rodger}.}
  \bibinfo{year}{2006}\natexlab{}.
\newblock \bibinfo{booktitle}{\emph{{JFLAP: An Interactive Formal Languages and
  Automata Package}}}.
\newblock \bibinfo{publisher}{Jones and Bartlett Publishers, Inc.},
  \bibinfo{address}{USA}.
\newblock
\showISBNx{0763738344}


\bibitem[Rodger et~al\mbox{.}(2006)]%
        {RodgerII}
\bibfield{author}{\bibinfo{person}{Susan~H. Rodger}, \bibinfo{person}{Bart
  Bressler}, \bibinfo{person}{Thomas Finley}, {and} \bibinfo{person}{Stephen
  Reading}.} \bibinfo{year}{2006}\natexlab{}.
\newblock \showarticletitle{{Turning automata theory into a hands-on course}}.
  In \bibinfo{booktitle}{\emph{Proceedings of the 37th {SIGCSE} Technical
  Symposium on Computer Science Education, {SIGCSE} 2006, Houston, Texas, USA,
  March 3-5, 2006}}, \bibfield{editor}{\bibinfo{person}{Doug Baldwin},
  \bibinfo{person}{Paul~T. Tymann}, \bibinfo{person}{Susan~M. Haller}, {and}
  \bibinfo{person}{Ingrid Russell}} (Eds.). \bibinfo{publisher}{{ACM}},
  \bibinfo{pages}{379--383}.
\newblock
\urldef\tempurl%
\url{https://doi.org/10.1145/1121341.1121459}
\showDOI{\tempurl}


\bibitem[Sipser(2013)]%
        {Sipser}
\bibfield{author}{\bibinfo{person}{Michael Sipser}.}
  \bibinfo{year}{2013}\natexlab{}.
\newblock \bibinfo{booktitle}{\emph{Introduction to the Theory of Computation}
  (\bibinfo{edition}{3rd} ed.)}.
\newblock \bibinfo{publisher}{Cengage Learning}.
\newblock
\showISBNx{9781133187790}


\bibitem[Venables and Haywood(2003)]%
        {Venables}
\bibfield{author}{\bibinfo{person}{Anne Venables} {and} \bibinfo{person}{Liz
  Haywood}.} \bibinfo{year}{2003}\natexlab{}.
\newblock \showarticletitle{{Programming Students NEED Instant Feedback!}}. In
  \bibinfo{booktitle}{\emph{Proceedings of the Fifth Australasian Conference on
  Computing Education - Volume 20}} (Adelaide, Australia)
  \emph{(\bibinfo{series}{ACE '03})}. \bibinfo{publisher}{Australian Computer
  Society, Inc.}, \bibinfo{address}{AUS}, \bibinfo{pages}{267–272}.
\newblock
\showISBNx{0909925984}


\bibitem[Walter et~al\mbox{.}(2009)]%
        {Walter}
\bibfield{author}{\bibinfo{person}{Tobias Walter}, \bibinfo{person}{Fernando
  Silva~Parreiras}, {and} \bibinfo{person}{Steffen Staab}.}
  \bibinfo{year}{2009}\natexlab{}.
\newblock \showarticletitle{OntoDSL: An Ontology-Based Framework for
  Domain-Specific Languages}. In \bibinfo{booktitle}{\emph{Model Driven
  Engineering Languages and Systems}}, \bibfield{editor}{\bibinfo{person}{Andy
  Sch{\"u}rr} {and} \bibinfo{person}{Bran Selic}} (Eds.).
  \bibinfo{publisher}{Springer Berlin Heidelberg}, \bibinfo{address}{Berlin,
  Heidelberg}, \bibinfo{pages}{408--422}.
\newblock
\showISBNx{978-3-642-04425-0}


\bibitem[White and Way(2006)]%
        {White}
\bibfield{author}{\bibinfo{person}{Timothy~M. White} {and}
  \bibinfo{person}{Thomas~P. Way}.} \bibinfo{year}{2006}\natexlab{}.
\newblock \showarticletitle{{jFAST: A Java Finite Automata Simulator}}.
\newblock \bibinfo{journal}{\emph{SIGCSE Bull.}} \bibinfo{volume}{38},
  \bibinfo{number}{1} (\bibinfo{date}{March} \bibinfo{year}{2006}),
  \bibinfo{pages}{384--388}.
\newblock
\showISSN{0097-8418}
\urldef\tempurl%
\url{https://doi.org/10.1145/1124706.1121460}
\showDOI{\tempurl}


\end{thebibliography}


\end{document}